# Using Economic Risk to Model Miner Hash Rate Allocation in Cryptocurrencies

George Bissias[1], Brian N. Levine[1], and David Thibodeau[2]

[1] College of Information and Computer Sciences, UMass Amherst
{gbiss,levine}@cs.umass.edu
[2] Tampa, Florida
davidpthibodeau@gmail.com

**Abstract.** Abrupt changes in the miner hash rate applied to a proof-of-work (PoW) blockchain can adversely affect user experience and security. Because different PoW blockchains often share hashing algorithms, miners face a complex choice in deciding how to allocate their hash power among chains. We present an economic model that leverages Modern Portfolio Theory to predict a miner's allocation over time using price data and inferred risk tolerance. The model matches actual allocations with mean absolute error within 20% for four out of the top five miners active on both Bitcoin (BTC) and Bitcoin Cash (BCH) blockchains. A model of aggregate allocation across those four miners shows excellent agreement in magnitude with the actual aggregate as well a correlation coefficient of 0.649. The accuracy of the aggregate allocation model is also sufficient to explain major historical changes in inter-block time (IBT) for BCH. Because estimates of miner risk are not time-dependent and our model is otherwise price-driven, we are able to use it to anticipate the effect of a major price shock on hash allocation and IBT in the BCH blockchain. Using a Monte Carlo simulation, we show that, despite mitigation by the new difficulty adjustment algorithm, a price drop of 50% could increase the IBT by 50% for at least a day, with a peak delay of 100%.

## 1 Introduction

Understanding how and why miners apply their hash rate to a given proof-of-work (PoW) blockchain is critical to understanding both the security and user experience of that chain. *In this paper, we show that miner hash rate allocations among blockchains can be largely explained by miner risk tolerance and fiat trade price movements in the coins minted by those chains.* Thus, abrupt changes in the price of one coin relative to the others can cause an abrupt change in miner hash rate allocations. A rapid drop in hash rate on a given blockchain presents a security risk in that the probability of a double-spend attack increases inversely proportional to the work applied to the chain [15]. A sudden drop in hash rate can also result in a temporary increase in inter-block time, which constitutes a lapse in user experience. Such a direct link between trade price and service quality is without precedent among conventional financial services. It is analogous to credit card transactions being processed more slowly whenever the stock price of Visa Inc. drops.

Miners typically invest in ASICs, which are hardware implementations of a particular PoW algorithm. Therefore, they can easily shift or *reallocate* their hash rate between blockchains that share the same PoW algorithm. Currently, the two largest blockchains, by market cap, that share the same algorithm are Bitcoin (BTC) and Bitcoin Cash (BCH). It is broadly acknowledged that the price of BCH relative to BTC is a strong determinant of miner allocation [3–5]. But direct comparisons of prices is problematic. For example, the *difficulty adjusted reward index* (DARI) is a popular measure of the relative profitability of mining on BTC versus BCH [6]. However, according to the DARI, one coin is always more profitable than the other — so why does each miner typically divide its hash rate allocations between chains? In the present work, we show how the allocation can be explained by miners' tolerance to changes in coin prices.

**Contributions.** Our primary contribution is an economic model for miner hash rate allocation, which we develop as an application of the Modern Portfolio Theory of Markowitz [13]. We show that the model is capable of accurately explaining the hash rate allocations of four out of the top five mining pools mining both BTC and BCH over the past six and a half months[3]. Over the entire timespan, the model's mean absolute error is at or less than 20% for those four miners, and the predicted aggregate allocation demonstrates a Pearson correlation coefficient of 0.649 when compared to actual. In contrast, estimates of allocations based only on short-term price changes or the DARI result in correlation coefficients of just 0.298 and 0.165, respectively. Our second contribution is demonstrating that the hash rate allocation resulting from the economic model is capable of accurately predicting major changes in the inter-block time (IBT) for BCH. Over the same time interval, the IBT predicted by our price model shows a Pearson correlation coefficient of 0.849 with the actual IBT. Because our predictions are based primarily on historical prices, the implication of this strong agreement is that deviation in IBT can be largely explained by fluctuations in coin prices. Finally, we use synthetic price data and hash rate allocations from our economic model in simulation to shows how the IBT would be affected by large fluctuations in the price of BCH. We find that even with the new difficulty adjustment algorithm employed by BCH, a price drop of 50% could increase the IBT by 50% for at least a day, with a peak delay of 100%.

## 2 Background

**Mining markets.** Nearly every blockchain project uses a proof-of-work (PoW) algorithm that is common to other projects. For example, of the top 50 cryptocurrencies by market capitalization[4], eight use SHA256 [15] including Bitcoin Cash and Bitcoin, seven use Ethash [2] including Ethereum and Ethereum Classic, and 11 use Scrypt [16] including Litecoin and Dogecoin. For PoW algorithms common to multiple currencies, miners are able to apportion their hardware among them, and can also rapidly change this allocation. Miners began manufacturing ASICs for SHA256 in 2013 [7], and ASICs for the Scrypt algorithm became available in 2014 [1]. ASICs for the Ethash algorithm were also introduced recently [20]. Mining with ASICs requires a large capital expenditure to purchase the hardware, and that investment has the effect of locking miners into a specific PoW algorithm in the medium-term. As a result, blockchain projects that share the same PoW algorithm form *multi-chain mining markets* comprised of miners who possess ASICs suitable for that algorithm.

[3] We will make our data publicly available before publication.
[4] https://minethecoin.com

**Profitability of mining.** Although mining markets have existed for several years, a miner's choice of hash rate distribution among blockchains, which we term *allocation*, remains somewhat mysterious [8]. A complicating factor is that most miners participate via a *mining pool*, which aggregates the hash power of its constituents and distributes mining rewards according to their hash rate. The allocation represented by a mining pool depends on that pool's policy. Most allow individual miners to either choose the blockchain on which they wish to mine, or follow the pool's choice of the most profitable chain [4,5]. However, it's not always obvious how to determine profitability [3]. What seems clear is that the choice of allocation is related to short-term profitability [6]. But long-term financial and idealogical concerns likely also play an important role.

Ignoring idealogical and long-term financial determinants, there are several factors that contribute to the relative profitability of mining between blockchains in the same mining market, including: *(i)* the relative fiat trade value of each coin; *(ii)* any hinderances to converting mining profits into fiat currency (e.g., poor coin liquidity); and *(iii)* the relative difficulty in mining the coins. The question of relative difficulty is particularly interesting from a technical standpoint because generally each blockchain in a mining market implements a different difficulty adjustment algorithm (DAA). The update frequency and accuracy of each DAA, relative the others, plays a critical role in how profitability changes over time.

**Difficulty adjustment algorithms.** In this paper, we present an in-depth analysis of mining profitability in the SHA256 mining market where Bitcoin (BTC) and Bitcoin Cash (BCH) together comprise 99% of the market cap; together these two comprise 67% of the market cap of all cryptocurrencies. In BTC, the difficulty is recalibrated every 2016 blocks by adjusting it either up or down inversely proportional to the deviation in mean inter-block time from optimal[5]. Since the hard fork on November 13, 2017, BCH performs a similar adjustment except that it occurs every block and covers a window of 144 prior blocks[6]. Prior to the November 13 hard fork, BCH used the same DAA as BTC except that it also implemented an Emergency Difficulty Adjustment algorithm (EDA) [18]. The EDA simply cut the difficulty by 20% any time that it took more than 12 hours to mine the last six blocks.

[5] http://github.com/bitcoin/bitcoin/blob/master/src/pow.cpp#L49

[6] http://github.com/bitcoincashorg/bitcoincash.org/blob/master/spec/nov-13-hardfork-spec.md

## 3 Related Work

There are several past works related to our contributions. To the best of our knowledge, we are the first work to evaluate, in a multi-blockchain market, the link between prices, hash rate allocation, and system performance. Most past work related to mining efficiency has focused on mining on a single blockchain. Rosenfeld [17] was one of the first authors to explore financial incentives in mining pools. He detailed several payout schemes and showed how they fair against several types of miner behavior. One particularly interesting behavior is called *pool hopping*, which involves a miner switching between pools mining the same coin in order to gain higher profits. This behavior is the *single-blockchain* analog to the *multi-blockchain* mining we analyze in this paper. Fisch et al. [12] conducted an analysis of pool payout strategies for mining on a single blockchain using discounted utility theory. They found that the geometric pay pool — in which rewards are concentrated at the winning block and decay exponentially over the preceding shares — achieves the optimal equilibrium utility for miners. Our focus is not on payout strategies for pools.

Meshkov et al. [14] considered miners switching between multiple blockchains. They argued that it is profitable for a miner to *hop* between blockchains with the same PoW algorithm, causing oscillations in difficulty that the miner can use to boost profit. The paper calculates the expected additional reward for the miner and shows that under this scheme the expected average inter-block time (IBT) on both chains exceeds the target time. The work is similar to ours in that it considers the profitability of moving hash rate between blockchains — however, it stops short of developing an economic model of hash rate allocation. In particular, they do not account for the influence of coin price on allocation; nor do they attempt to determine an equilibrium allocation. Moreover, it is not clear that chain hopping is currently pervasive in blockchains. For example, if miners do commonly engage in chain hopping on BCH, then their theory anticipates that IBT should substantially exceed the target of 600 seconds, but we find the mean IBT to be 604 seconds since the November 13 hard fork.

Several authors have formulated economic models of the mining ecosystem in an effort to predict or explain coin price. In contrast, we are not attempting to discover what drives price, but rather how price drives system performance. For example, Cocco and Marchesi [10] used an agent-based model of the mining process to show its relationship to Bitcoin price. The model had some success in predicting large price peaks as well as some statistical properties of the Bitcoin ecosystem. Chiu et al. [9] developed a general equilibrium monetary model for Bitcoin and similar cryptocurrencies. A major consequence of the model is that cryptocurrencies must trade off between immediacy and finality of settlements.

## 4 Miner Hash Allocation

In this section, we develop an theory of how and why, in economic terms, miners distribute their hash power among competing blockchains. The recent split of

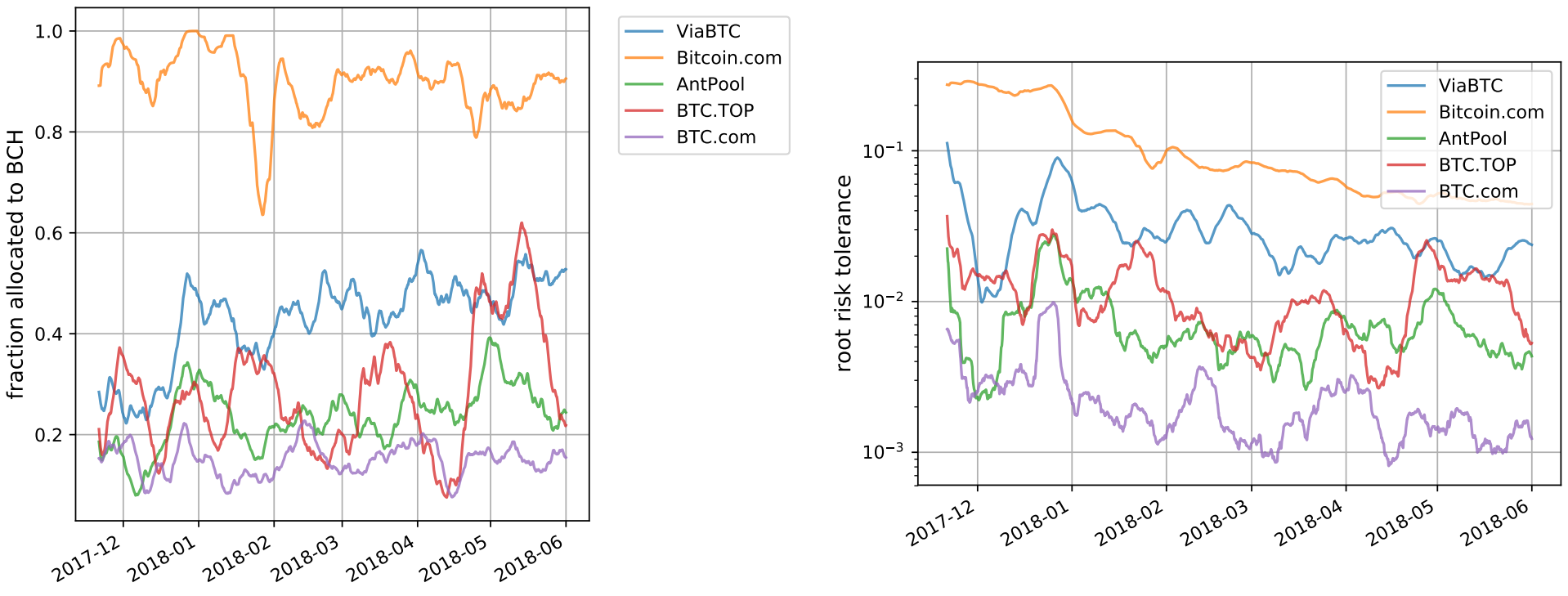

Fig. 1: Fraction of miner hash rate allocated to the BCH blockchain instead of the BTC blockchain (left) and square root of risk tolerance (right). Shown here are only the top five miners that historically mine on both chains. In the right plot, risk tolerance is in units of the USD price of BTC + BCH at the given time.

Bitcoin Cash (BCH) from Bitcoin (BTC) provides an important case study for us: each currency is highly valued and both rely on the same PoW algorithm. As a result, it is trivial for miners to distribute their hash power among the two blockchains as they see fit. This presents a conundrum for us: at any given time, it is almost always more profitable to mine exclusively on one chain or the other; yet, among miners that participate in mining on both chains, hash rate allocation is typically divided between the two. Figure 1 (left) plots the history of several mining pools' allocation of hash rates to BCH as a fraction of their respective total resources. Thus, it appears that miners are not actually acting in purely greedy fashion, and we require a model that accounts for this nuance.

*We hypothesize that miners are acting in a manner so as to maximize their profit, subject to a particular risk tolerance.* That is to say, miners seek greater profits, but they are also sensitive to the high volatility of holding cryptocurrencies. The exposure to this volatility is quantifiable: most blockchains impose a multi-block *cooldown period* during which miners are not allowed to spend their newly minted coins. For example, in both BTC and BCH, miners are required to hold their mined coins for a minimum of 101 blocks (roughly 17 hours)[7]. Thus, miners hold a short-term *portfolio* of the cryptocurrencies they mine. The Modern Portfolio Theory (MPT) of Markowitz [13], a seminal result in economics, provides a framework for determining the best allocation of assets with respect to profit expected value and volatility. We next develop a model of optimal miner hash rate allocation using the MPT framework.

### 4.1 An Economic Model

Consider a set of distinct blockchains $\boldsymbol{C} = [C_1, \ldots, C_n]$ that share the same PoW algorithm, and let vector $\boldsymbol{\pi}$ denote the miner profit for each. For miner $j$, define

[7] http://github.com/bitcoin/bitcoin/blob/master/src/consensus/consensus.h#L19, http://github.com/BitcoinUnlimited/BitcoinUnlimited/blob/release/src/consensus/consensus.h#L31

$\boldsymbol{w_j} = [w_{1j}, \ldots, w_{nj}]$ to be the *allocation* of this hash rate to the blockchains $\boldsymbol{C}$. And let vector $\boldsymbol{h} = [h_1, \ldots, h_m]$ denote the total hash rate for each miner across all blockchains. The *aggregate allocation* among all miners is given by

$$\boldsymbol{w} = \sum_j \boldsymbol{w}_j \frac{h_j}{\boldsymbol{e}^T \boldsymbol{h}}, \tag{1}$$

where $\boldsymbol{e}$ is the vector of all ones. Aggregate allocation captures the overall distribution of mining power among all blocks chains $\boldsymbol{C}$. Next, define $\Sigma = Cov(\boldsymbol{\pi})$, or the covariance of $\boldsymbol{\pi}$, which we call the *volatility matrix*. For a miner's allocation $\boldsymbol{w_j}$, the *risk* is given by $\boldsymbol{w_j}^T \Sigma \boldsymbol{w_j}$. And the *risk tolerance* of miner $j$, given by $\rho_j$, is defined as his maximum allowable risk. MPT predicts that a rational miner $j$ seeking to maximize expected profits will solve the following problem (although perhaps not explicitly):

**PROBLEM** `MaxProfit`$(j)$**:**

*Maximize*: $E[\boldsymbol{w_j}^T \boldsymbol{\pi}]$
*Subject to:* $\boldsymbol{w_j}^T \Sigma \boldsymbol{w_j} = \rho_j$, $\boldsymbol{w_j}^T \boldsymbol{e} = 1$,
and $\boldsymbol{e} = [1, \ldots, 1]$

We solve `MaxProfit` using Lagrange multipliers in similar fashion to Dhrymes [11]. However, in our formulation we do not allow for a portion of the portfolio to be allocated at the risk-free rate of return because we assume that miners are locked into their investment in mining hardware. Thus, we solve the system of equations associated with the critical points of the following Lagrangian:

$$\begin{aligned} L_j = {} & \boldsymbol{w_j}^T E[\boldsymbol{\pi}] + \\ & \lambda_{j1}(\rho_j - \boldsymbol{w_j}^T \Sigma \boldsymbol{w_j}) + \\ & \lambda_{j2}(1 - \boldsymbol{w_j}^T \boldsymbol{e}), \end{aligned} \tag{2}$$

which yields the following solution.

$$\begin{aligned} \boldsymbol{w_j} &= \Sigma^{-1} \tfrac{E[\boldsymbol{\pi}] - \lambda_{j2} \boldsymbol{e}}{2\lambda_{j1}} & a &= e^T \Sigma^{-1} e \\ \lambda_{j1} &= \tfrac{1}{2}(b - a\lambda_{j2}) & b &= e^T \Sigma^{-1} E[\boldsymbol{\pi}] \\ \lambda_{j2} &= \tfrac{b}{a} \pm \tfrac{\sqrt{(b^2 - ac)(1 - a\rho_j)}}{a(1 - a\rho_j)} & c &= E[\boldsymbol{\pi}]^T \Sigma^{-1} E[\boldsymbol{\pi}]. \end{aligned} \tag{3}$$

### 4.2 Profit and volatility in multi-chain mining

In a typical portfolio optimization problem [13], the profit for an asset, $\boldsymbol{\pi}$, is defined as the change in asset value over a given period of time $\Delta t$. However, miners are *creating* assets as opposed to merely acquiring them, so their profit should nominally account for the full fiat trade value of each coin that they mine. Still, miners contribute hash power to each blockchain, which amounts to an associated cost. Therefore, the ideal measure of profit is one that normalizes the fiat price of cumulative coinbase rewards by the relative difficulty.

Another complication is that miners can change their allocation at any time and for little-to-no cost. Thus, we hypothesize that they will re-evaluate Problem `MaxProfit` at every instant $t$. Hence, we seek parameterized representations of the profit vector and volatility matrix: $\boldsymbol{\pi}(t)$ and $\Sigma(t)$. To that end, let $\boldsymbol{R} = [R_1(t), \ldots, R_n(t)]$ be a vector representing the fiat value of coinbase reward for each blockchain at time $t$ (fees are ignored in this model). And define $\boldsymbol{D} = [D_1(t), \ldots, D_n(t)]$ to be the associated difficulties for those chains at the same time. We define the profit at time $t$ by

$$\boldsymbol{\pi}(t) = \boldsymbol{R}(t)/\boldsymbol{D}(t)\frac{\boldsymbol{e}^T\boldsymbol{D}(t)}{\boldsymbol{e}^T\boldsymbol{R}(t)}, \tag{4}$$

where "/" denotes component-wise division and $\boldsymbol{e}$ is the vector of all ones. Note that our definition for $\boldsymbol{\pi}(t)$ is equivalent to the *Difficulty Adjusted Reward Index* (DARI), a popular mining profitability metric [6], except that we ignore fees and normalize by the aggregate fiat value of all chains, $\boldsymbol{e}^T\boldsymbol{R}(t)$, and total difficulty, $\boldsymbol{e}^T\boldsymbol{D}(t)$. Normalizing by $\boldsymbol{e}^T\boldsymbol{R}(t)$ is necessary because cryptocurrency prices can fluctuate significantly over short periods, and normalization allows us to more directly compare profits at different times. Similarly, normalizing by $\boldsymbol{e}^T\boldsymbol{D}(t)$ allows us to ignore the effect of fluctuations in total hash rate on miner profit.

Given our definition for $\boldsymbol{\pi}(t)$, the expected profit vector, $E[\boldsymbol{\pi}(t)]$, can be approximated by the sample mean over all $\boldsymbol{\pi}$ from time $(t - \Delta t)$ until $t$. For volatility, we hypothesize that miners are concerned about price changes only over the short cooldown period $\Delta c$ that extends from the time a coin is mined until the time it can be traded for fiat currency. Thus, we seek to capture relative changes in profit between all blockchains during $\Delta c$. For simplicity, we assume that $\Delta c$ is the same for all chains. Finally, we define the volatility matrix by $\Sigma(t) = Cov(\boldsymbol{\pi}(t) - \boldsymbol{\pi}(t - \Delta c))$. $\Sigma(t)$ can be approximated by the sample covariance over the set of vectors: $\{\boldsymbol{\pi}(x) - \boldsymbol{\pi}(x - \Delta c) \mid t - \Delta t \leq x \leq t\}$.

## 5 Model Validation and Parameter Fitting

In general, a miner's choice in hash rate allocation results from a complex combination of economically rational profit seeking and more subtle idealogical considerations. As such, we do not expect that the solution to Problem `MaxProfit` can fully predict miner allocations; however, in this section we seek to demonstrate that it is capable of explaining much of their behavior. To do so, we analyzed approximately 6.5 months of price and blockchain data from BCH and BTC between November 14, 2017 and June 1, 2018. We intentionally omit data prior to the BCH hard fork on November 13, 2017, which introduced a new DAA. Prior to the fork, both BCH block times and prices were exceptionally irregular due to high price volatility as well as rampant manipulation of the EDA [19]. As a result, it is very difficult to accurately estimate actual miner allocations or infer their risk tolerance during the EDA time period.

For each blockchain, we calculated the time-parameterized profit vector and volatility matrix as described in Section 4.2 using hourly price data from the Bitfinex exchange. We chose unique but fixed values for lookback $\Delta t$ and risk tolerance $\rho$ for each miner using the techniques described in Sections 5.1 and 5.2. For both BTC and BCH chains, we set $\Delta c = \lfloor 101/6 \rfloor$ hours to match their 101-block cooldown period. We analyzed each of the top five mining pools on BCH that are also active on BTC, excluding the mining by pools that do not claim blocks. We determined the actual allocations for each miner, $\boldsymbol{w}_j(t)$, by first calculating the average fraction of blocks produced per hour on each blockchain using an exponentially weighted moving average with a half-life of 10 hours. Estimating hash allocation from mined blocks is very noisy, and using a weighted average of recent blocks allowed us to arrive at a more smooth estimate. These average block rates were translated into allocations after normalizing by the relative difficulty of each chain.

### 5.1 Inferred Miner Risk

Our economic model predicts that each miner allocates her hash rate based on the historical profit for each coin as well as her personal risk tolerance. A miner's risk can be inferred from her current allocation and volatility matrix $\Sigma(t)$. According to the model, we assume that any given miner will exhibit a consistent risk tolerance. Furthermore, for a given risk tolerance, we anticipate that the actual allocation chosen by miners will match the economic allocation produced by the model.

Risk $\rho_j$ is measured in units of squared deviation in profit. And because profit is normalized by the sum of fiat prices of each chain in $\boldsymbol{C}$ (see Section 4.2), the square root of risk, or *root risk*, also has units of BTC + BCH (which we write as BTC+, for short). Therefore, the root risk can be interpreted as the maximum deviation in profit, in units of BTC+, that is tolerated by the miner during the cooldown period $\Delta c$. For example, when 1 BTC trades for 10 BCH (BCH/BTC = 0.1), a miner with root risk 0.043, who is allocated entirely to BCH, will tolerate a decrease to BCH/BTC = 0.05 during $\Delta c$.

Figure 1 (right) shows the root risk for each of the top five mining pools that mine both BTC and BCH. The relative risk tolerance among miners remains very consistent over time. The Bitcoin.com mining pool exhibits the highest risk tolerance, while BTC.com shows the lowest. ViaBTC maintains root risk roughly between 0.01 and 0.1 BTC+, while AntPool and BTC.TOP typically range from 0.003 to 0.03 BTC+, and BTC.com fluctuates between 0.001 and 0.01 BTC+. In absolute terms, Bitcoin.com also demonstrates largest variation in risk tolerance, showing a high of 0.3 BTC+ at the end of November and recent low near 0.06 BTC+. From Figure 1 (left) we can see that differences in risk tolerances are roughly reflected by the choice in miner allocations. For example, Bitcoin.com is mostly allocated to mining BCH, while BTC.com mines BTC almost exclusively.

Figure 2 juxtaposes the risk associated with mining exclusively on the BTC or BCH blockchains with the BCH/BTC trade price ratio taken from USD quotes on the Bitfinex exchange. The risks for each blockchain were calculated using a lookback of $\Delta t = 48$. Problem `MaxProfit` utilizes information from all three facets to derive the economic allocation. There are several notable features in these curves. First, from Figure 1 (right), we can see that risk rose sharply for miners allocated to BCH near the end of 2017. The top two facets of Figure 2 indicate that this was a period where risk in mining BCH rose far faster than for BTC, while the bottom facets shows that BCH simultaneously made major gains on BTC in terms of price. We hypothesize that this indicates that miners are willing to relax their risk tolerance at times when they anticipate major gains for one coin over another (in this case BCH over BTC). Second, not all major price movements will result in increased risk for the current allocation. Because it was gradual, the increase in the price of BCH relative to BTC at the end of April is not accompanied by a large rise in risk for either blockchain. Nevertheless, we do see from Figure 1 (left) that mining pools BTC.TOP and AntPool substantially increased their BCH allocation during this time. As a result, their risk rose accordingly.

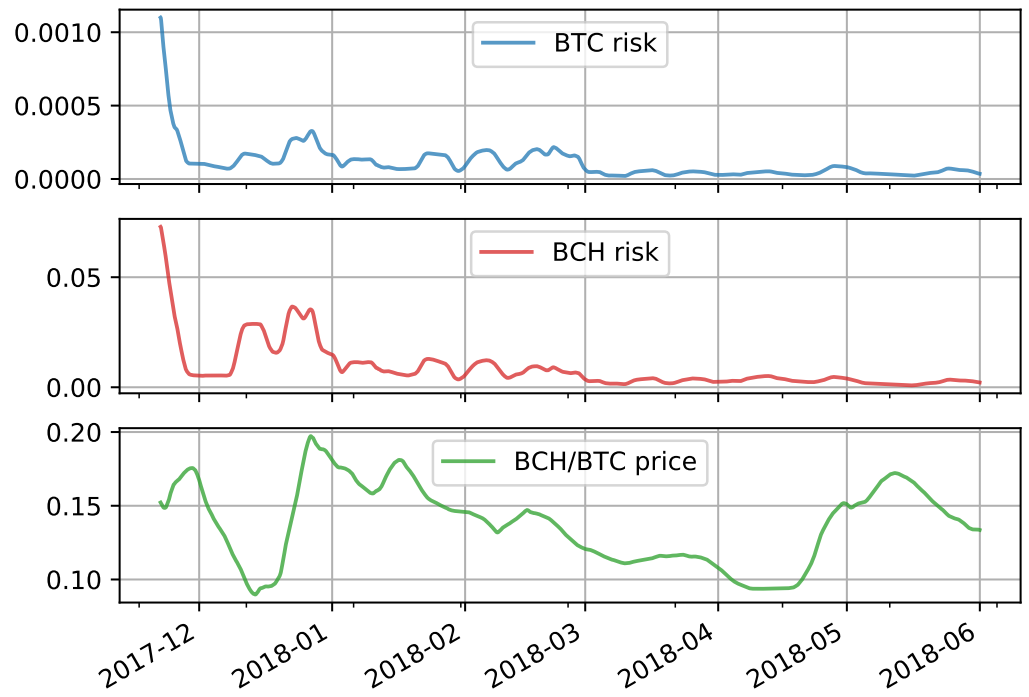

Fig. 2: Risk and price juxtaposed. The top two facets show the risk for BTC and BCH associated with allocating all hash rate to either the BTC or BCH blockchain, respectively. The bottom facet shows the price ratio of BCH to BTC; the price for each was drawn from the Bitfinex exchange where it was quoted in terms of USD.

| **Mining Pool** | **Lookback** | **Risk** | **Mean error** |
|---|---|---|---|
| ViaBTC | 144 | 6.42 e-04 | 20.0% |
| BTC.TOP | 16 | 8.54 e-05 | 20.7% |
| Bitcoin.com | 1,008 | 2.40 e-03 | 36.0% |
| AntPool | 10 | 3.33 e-05 | 17.0% |
| BTC.com | 4 | 3.81 e-06 | 14.4% |

Table 1: Optimal lookback (hours) and risk parameters and mean absolute error for the top 5 miners who mine both BTC and BCH based on observable data.

Despite the tendency for risk tolerance to fluctuate during abrupt price movements, Figure 1 (right) still reflects overall consistency in inferred root risk for most miners except Bitcoin.com. For the remaining miners, we believe that a single risk tolerance $\rho_j$ chosen for each miner $j$ is sufficient to describe much of that miner's behavior, and therefore our economic model may provide a reliable prediction of their allocation of hash power. In order to choose $\rho_j$ for a given miner, we tested 8 equally spaced risk values falling between the 25th and 75th percentiles of the historical inferred risk values for that miner. For each risk value, and each possible lookback chosen from the set described in Section 5.2, we calculated the economic allocations using our model and compared them to the actual allocations chosen by the miner using the Kolmogorov-Smirnov test for goodness-of-fit. We selected the value for $\rho_j$ that yielded the best fit of the economic allocation to the actual. Results are shown in Table 1.

### 5.2 Determining miner lookback period $\Delta t$

Risk is only one factor used to determine the optimal allocation. Another important factor is the lookback period $\Delta t$. This period dictates how much historical data will be used to calculate expected profit and volatility. For miners there is a tradeoff between accuracy and immediacy. On one hand, using the entirety of historical data will yield the most accurate estimate of the overall value of the statistics. But on the other hand, older data is likely to be less relevant, particularly when market characteristics can change abruptly.

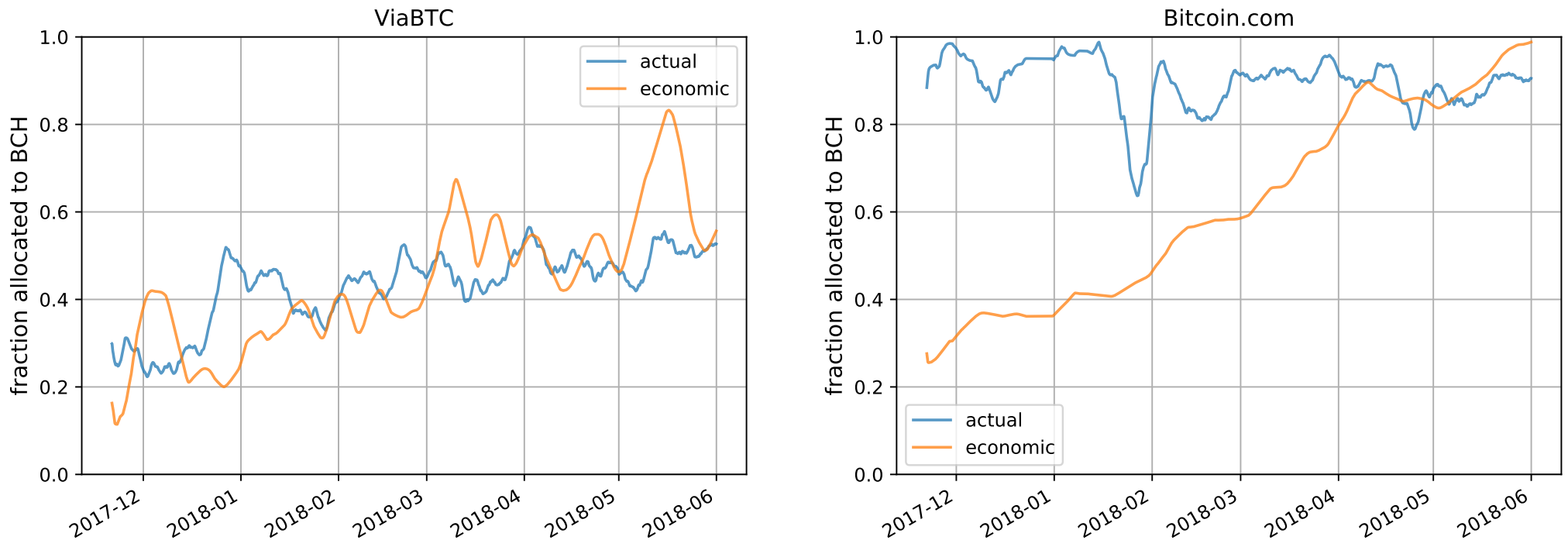

Fig. 3: Actual and economic hash rate allocations for the two largest pools that mine both BTC and BCH: ViaBTC (left) and Bitcoin.com (right). In each figure, the actual allocation (blue) is compared to the optimal economic allocation (orange), the latter of which is formed using parameters chosen from Table 1 for each miner.

We determined the optimal $\Delta t_j$ for miner $j$ by testing the following values.

$$S = \{4\text{–}24 \text{ in increments of } 6\} + \{24\text{–}144 \text{ in increments of } 24\} + \{168\text{–}1344 \text{ in increments of } 168\} \quad (5)$$

For each $\Delta t_j \in S$ and each potential risk value $\rho_j$ (chosen according to the procedure described in Section 5.1), we determined the optimal economic allocation by solving `MaxProfit` using statistics $E[\boldsymbol{\pi}(t)]$ and $\Sigma(t)$, which were formed as described in Section 4.2. We then chose the values for $\Delta t_j$ and $\rho_j$ for miner $j$ corresponding to the economic allocation that yielded the best fit relative to the actual allocation according to the Kolmogorov-Smirnov test. Table 1 shows the chosen values for $\Delta t_j$ and $\rho_j$ for the top five miners. We use these values in the remainder of our analysis.

### 5.3 Comparing actual to optimal allocations

Figure 3 compares actual allocations versus allocations from our risk and price-driven economic model for the two largest pools participating in both BTC and BCH mining: ViaBTC and Bitcoin.com. We determined the optimal economic allocations by selecting the parameters from Table 1 and solving Problem `MaxProfit`. Figure 3 shows strong agreement between economic and actual allocations for ViaBTC. On the other hand, the economic allocation for Bitcoin.com shows very poor agreement with actual during the months prior to April, 2018. As a result, we hypothesize that there do not exist any single values for risk tolerance and lookback that can describe the hash rate allocation of Bitcoin.com over the entire time period. This hypothesis is corroborated by Figure 1 (right), which shows that inferred root risk has been decreasing rapidly since late November 2017. For this reason, we omit the Bitcoin.com mining pool from the remainder of our analysis, as its allocations are not described well by our economic model.

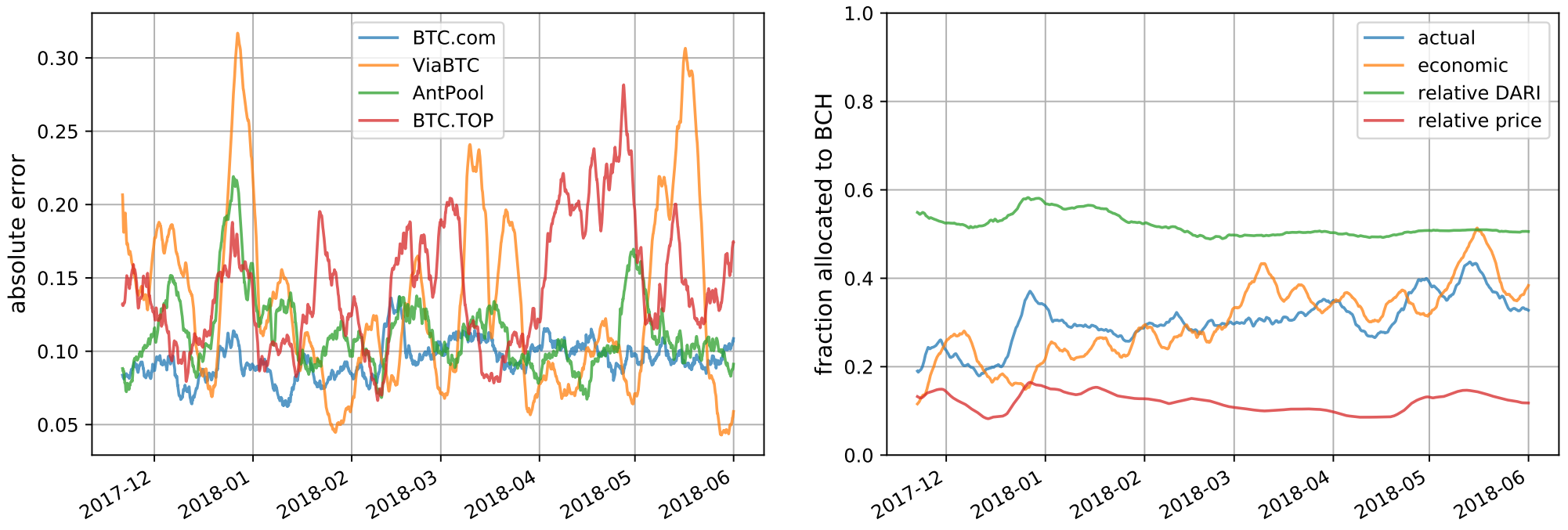

Fig. 4: Absolute error between economic and actual allocations (left) and aggregate allocation (right) for the top four pools (excluding Bitcoin.com) mining both BTC and BCH.

Figure 4 shows the absolute error and aggregate allocation for the top four pools (excluding Bitcoin.com) that participate in mining on both BTC and BCH. Together, these pools constitute approximately 48% of the total hash rate for BCH. From the plot of absolute error, we can see that economic and actual allocations are typically quite close for the four mining pools; Table 1 shows that their mean error is at or below 20%. The low error results in strong agreement between the actual and economic aggregate allocations, shown in Figure 4 (right). We used Equation 1 for aggregating both actual and economic allocations.

For comparison, we also plot two other price-driven allocations: $D(\text{BCH})$ / $(D(\text{BCH}) + D(\text{BTC}))$ and $P(\text{BCH})$ / $(P(\text{BCH}) + P(\text{BTC}))$. The function $D$ denotes the DARI, which is the value of the given chain's coinbase in USD divided by the current difficulty (we ignore fees). And the function $P$ denotes the USD trade price. Neither the relative DARI nor relative price show strong agreement with actual allocations. Their Pearson correlation coefficients are 0.165 and 0.298, respectively, and the magnitudes of the allocations are also quite different than actual. In contrast, the economic allocation provided by our model shows strong agreement with the actual allocations both in terms of correlation coefficient, 0.649, as well as general similarity in the magnitude of the allocation. For this reason we believe that it is valid to employ our economic model in describing the aggregate behavior of the top four mining pools, excluding Bitcoin.com.

## 6 Using Risk to Explain Change in Inter-Block Time

Based on the economic model introduced in Section 4.1, we hypothesize a direct relationship between short-term price fluctuations and deviation in inter-block time (IBT). In particular, we hypothesize that a large change in the expected profit $E[\boldsymbol{\pi}](t)$ will lead to a large change in a miner's hash rate allocation $\boldsymbol{w}_j(t)$, which will propagate to the aggregate allocation $\boldsymbol{w}(t)$ defined by Equation 1, and ultimately impact IBT until the difficulty is adjusted.

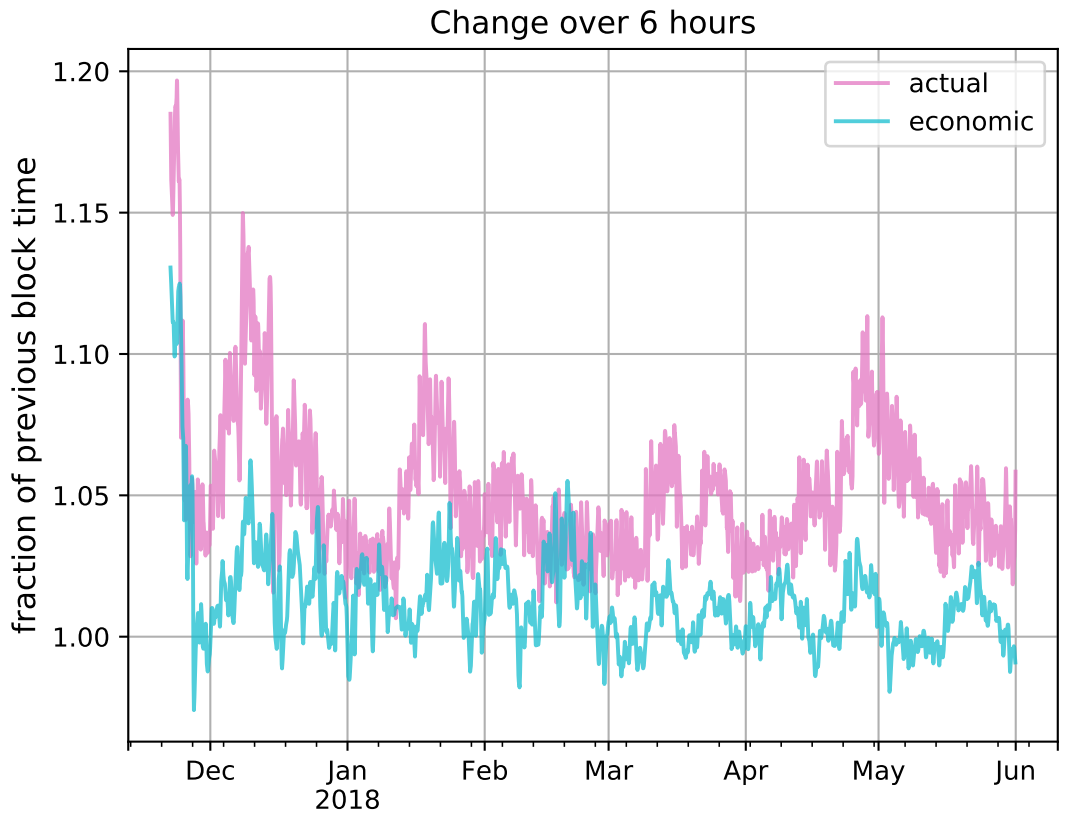

Fig. 5: Predicted (blue) and actual (pink) change in BCH inter-block time during one 6 hour period compared to the next (non-overlapping) 6 hour period. Predicted change in block time is calculated using Equation 1 and solving Problem `MaxProfit` with parameters chosen from Table 1.

Let $\boldsymbol{T} = [T_1, \ldots, T_n]$ denote the target IBT for each blockchain. If we assume that the elapsed time $\delta t$ was short enough that no blockchain has yet substantially updated its difficulty, then the expected IBT will have changed by

$$\delta \boldsymbol{T} = \boldsymbol{w}(t)/\boldsymbol{w}(t + \delta t) \circ \boldsymbol{T}, \tag{6}$$

where "$\circ$" and "/" denote element-wise vector multiplication and division.

**Prediction of change in IBT.** Equation 6 provides a means of using our economic model to predict the change in IBT from only historical price data and miner risk tolerances. We analyzed historical data from November 14, 2017, until June 1, 2018 using the aggregate economic allocation (with parameters chosen from Table 1) to estimate the change in IBT for the BCH blockchain from one 6 hour period to the next (non-overlapping) 6 hour period. The experiment used the top four mining pools, excluding Bitcoin.com, which constitute approximately 48% of the total hash rate on BCH during that time. Figure 5 shows the result of these predictions compared to actual change in IBT using a 7 day rolling average.

Despite being quite noisy, the figure shows a strong correlation between predicted and actual IBT change throughout the six and a half month timeframe. The Pearson correlation coefficient between predicted and actual IBT is 0.849. In addition to correlation, the predicted change in IBT also echoes the magnitude of changes in actual IBT. However, the predicted result does appear to consistently under-estimate the extent of change by as much as 0.05. There are two possible reasons for this inaccuracy. First, our price data is accurate only to the nearest hour, so it is possible that the full extent of large price shocks is not reflected in the economic allocation. And second, ignoring the effects of the DAA introduces a subtle bias. The DAA is much better at compensating for an IBT that is too short as opposed to too long. When the IBT is short, more blocks are arriving, so the algorithm has more opportunities to adjust the difficulty whereas no adjustments can be made between blocks. Thus, IBT change less than 1 tends to be minimal while change greater than 1 tends to be exaggerated. Indeed changes below 1 are small enough that the 7 day rolling average eliminates them entirely. But because the predicted IBT does not model the effects of the DAA, it treats drops in allocation identically to spikes.

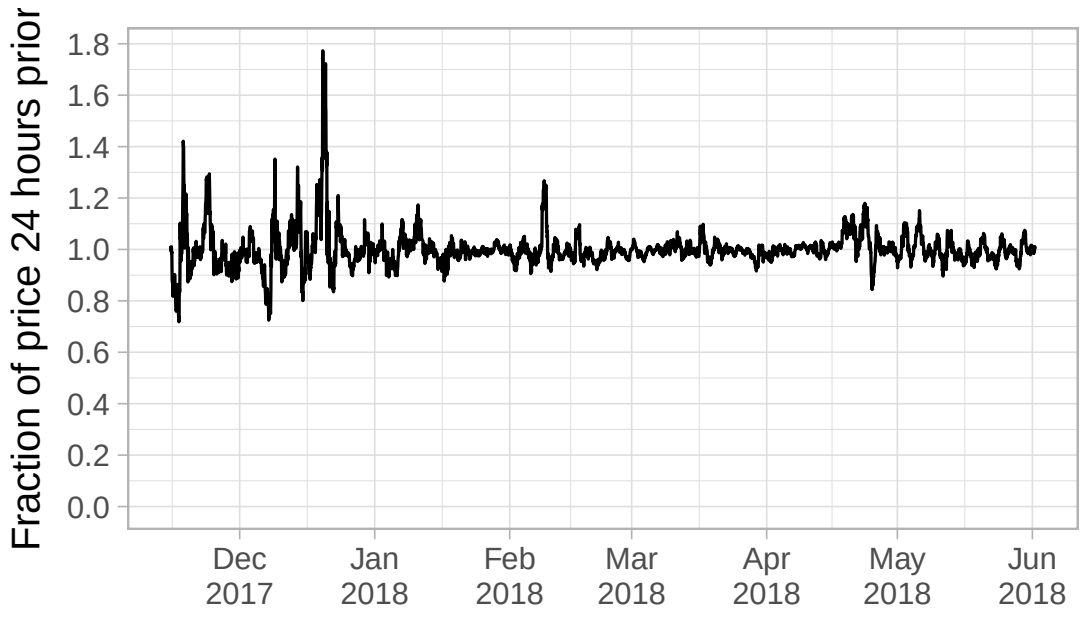

Fig. 6: The fraction of a given hour's price to the price 24 hours earlier, where price is defined as BCH/(BTC+BCH). 98% of the time, the fraction is between 0.8 and 1.2.

# 7 DAA Susceptibility to Price Shocks

In this section, we use our economic model to quantify how specific price changes affect inter-block times (IBTs) via changes in hash rate allocation. We show that even with a proactive controller that adjusts the difficulty every block, like the DAA currently implemented for BCH, large enough price shocks can still lead to long delays in IBT with affects being felt for a day or more. In reality, prices and their volatility are not the only determinants of miner behavior, but in Sections 5 and 6 we presented evidence that these economic factors are often sufficient for accurately explaining real world miner allocations and ultimately IBT.

Blockchains compensate for changes in hash rate with an algorithmic change in *difficulty*. Ideally, the difficulty is changed so that IBT remains at a desired mean, which is 600 seconds per block for BCH. Below, we quantify how a *price shock* — a single, sudden rise or drop in price of BCH compared to BTC — can change IBT given the current BCH difficulty adjustment algorithm (DAA). We begin by characterizing typical price changes in BCH relative to BTC using price data from November 2017 through May 2018. We then quantify how various changes in BCH price can affect allocation and IBT under a simplifying assumption that all miners are applying the economic model.

Figure 6 shows, for each hour, the fraction change in price from the previous 24 period to the next (non-overlapping) 24 hour period. Price is defined as BCH/(BTC+BCH). As the plot shows, 98% of the time, daily price changes are no greater than 20%. However, eight times the price changed approximately 30% or more and once it changed by 80%. Thus, there exists historical precedence for a *maximum* 24-hour change of nearly 100%.

Section 6 argued that the aggregate allocation given by solving Problem `MaxProfit` and applying Equation 1 can be used to roughly predict actual IBT changes even without taking into account the effect of the DAA on block time regularization. We speculated that our failure to take the effect of the DAA into account was a major cause of the downward bias in the prediction. Regardless of the reason for the bias, Figure 5 shows that the aggregate economic allocation can accurately predict major changes in IBT, and if anything, might tend to underestimate the extent of increases in IBT. Thus, we believe that our economic allocation provides a sufficiently accurate estimate of actual miner allocations to be used to predict the effect of a price shock on IBT.

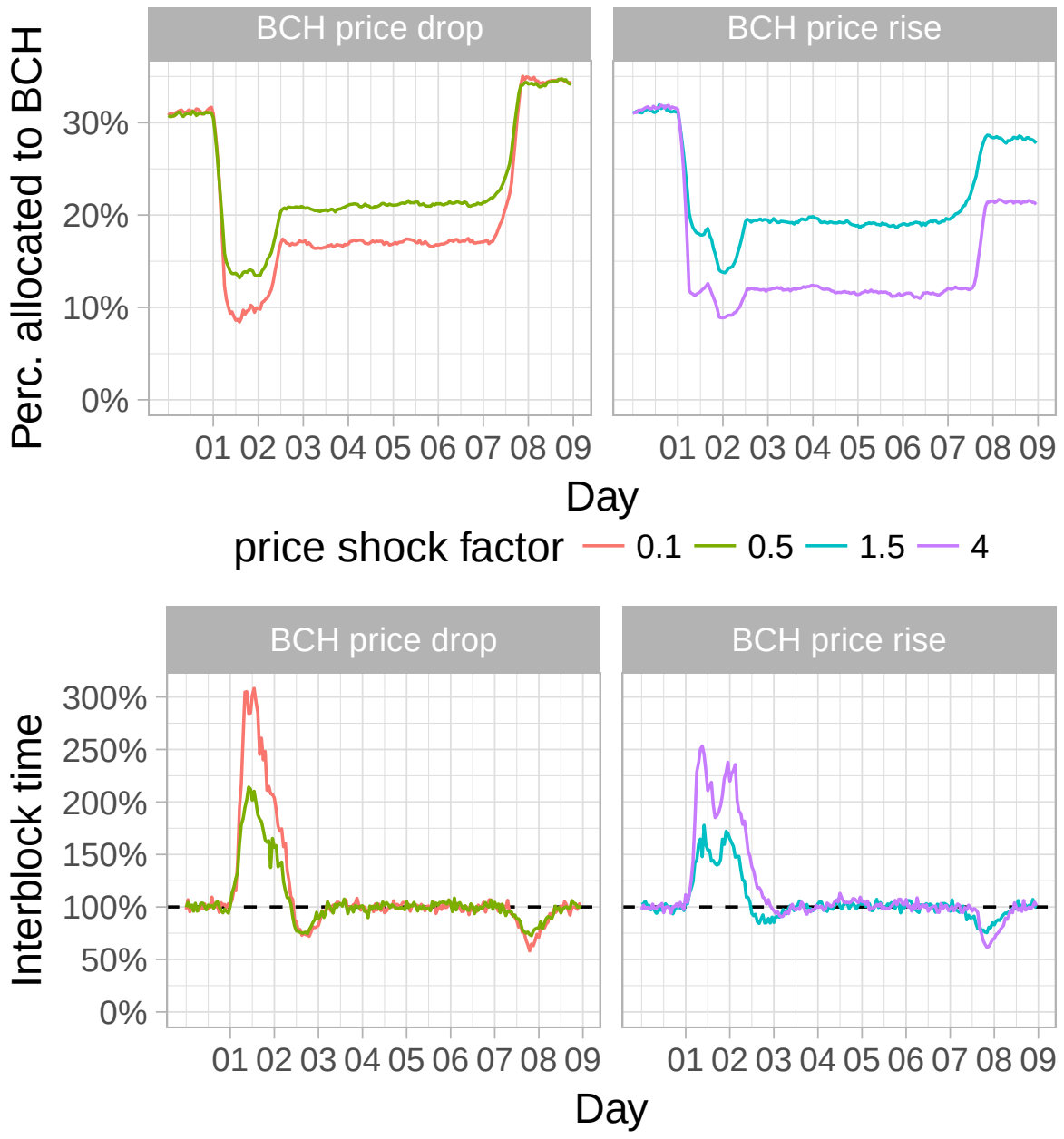

Fig. 7: The change in allocations (Top) and therefore inter-block time (Bottom) that results from a single price shock that is larger than typical. The price shock is determined by a multiplier, shown in the legend. Allocations decrease as volatility increases, which causes inter-block time to rise. For example a shock of 1.5 increases the price by 50% on day 1 and from there the price does not change. (From Monte Carlo simulation, risk and lookback parameters from the top four BCH miners excluding Bitcoin.com.)

**Price shock experiment.** To quantify the effects of various price shocks, we ran a block-generation simulator that updated the synthetic coin price every block. All prices for BCH were initially set to $p$, the mean USD value for BCH between November 2017 through May 2018. Each experiment introduced exactly one shock $x \in (0, 4]$, which set all prices subsequent to this *shock block* to $px$. Thus, the BCH prices for each experiment formed a step function with a step up in price after the shock block when $x > 1$ and a step down when $x < 1$. To establish baseline volatility, we also added uniform random noise in the range $[-0.1p, 0.1p]$ to all prices. Prices for BTC were generated similarly except that no shock was introduced and the base price $p$ was set to BTC's mean USD price over the same time period. For each experiment, corresponding to a separate shock, we ran at least 180 trials of the following Monte Carlo (MC) simulation. *(i)* We formed the aggregate economic allocation for the top four miners (excluding Bitcoin.com) by solving Problem `MaxProfit` using the synthetic prices for the given experiment and parameters from Table 1 and substituting the result into Equation 1. *(ii)* The difficulty was initially set to an arbitrary value and allowed to reach equilibrium at the pre-shock price. *(iii)* We stepped through the generation of each block, adjusting the allocation every block according to the economic allocations to determine the hash rate for the mining process. *(iv)* After each block, we ran the DAA to adjust the difficulty according to the IBT of the mined blocks.

Figure 7 (Top) shows the median change in economic allocation over all simulation trials that results from a single price shock $x$ given by the value shown in the legend. Figure 7 (Bottom) shows the corresponding changes in mean IBT. Overall, we see that, even with compensation from the DAA, a drop in price of as little as 50% can increase mean IBT by more than 50% for an entire day, while a drop to 10% of the original value can double the mean IBT for at least a day. Similarly, a rapid price increase by 50% is expected to raise the mean IBT by 50% for at least a day, and an 800% price increase could more than double the mean IBT for a day or more.

It is somewhat counterintuitive that both price drops (left plots) and increases (right plots) result in lower economic allocations initially, and in the long run, allocations actually stabilize to higher values after a price drop and lower values after a spike. Allocations drop immediately after the shock date because volatility has risen for BCH relative to BTC, regardless of the direction of the shock. Essentially the economic allocation follows the maxim, "what goes up must come down". However, it is reasonable to question how realistic this aspect of the economic model is during a price spike for BCH. Indeed, Figures 1 (left) and 2 indicate that all of the top five miners except Bitcoin.com increased their allocation in BCH after it massively gained in price on BTC at the end of 2017, despite the commensurate rise in risk. On the other hand, both ViaBTC and Bitcoin.com reduced their allocation after the price (and risk) increase at the end of May. The long-term rise in allocation after a price drop is simply due to the fact that the baseline volatility relative to BTC becomes slightly lower after prices have stabilized and cleared every miner's lookback period. The opposite is true for the relative volatility after a price spike.

Another feature of the price shock simulation is the delayed after-shock observed approximately seven days later. Mathematically, this is the result of the expiration of the longest lookback period, corresponding to ViaBTC (see Table 1). Prior to the date in question, there exist prices in the lookback from both before and after the shock. Thus, the volatility remains high relative to the baseline. However, once the last pre-shock price has cleared the lookback period, volatility abruptly returns to baseline, causing a substantial increase in allocation to BCH and a corresponding decrease in IBT. Over the course of approximately one day, the DAA returns the IBT to normal. Although we do believe that it is plausible that miners use price data from the recent past to determine their current allocation, it is perhaps unlikely that they implement such a hard cutoff as to produce a sudden shift in allocation. For that reason we regard the aftershock as a modeling idiosyncrasy.

## 8 Conclusions

We have presented an economic model of miner hash rate allocation inspired by Modern Portfolio Theory. The model is sufficient to explain, with low error, the individual allocations of four of the top five mining pools active on both BTC and BCH blockchains. Taken together, they form a very accurate model of aggregate miner allocation between BTC and BCH using only historical price data, a single risk value, and a single lookback period for each miner. Using this aggregate allocation alone, it is possible to correctly predict major changes in actual inter-block time (IBT). Our model is also capable of analyzing theoretical price scenarios. It predicts that either a 50% drop or increase in the price of BCH relative to BTC can increase BCH inter-block times by 50% for a day or more.